\documentclass[twocolumn,10pt,aps,showpacs,prc,preprintnumbers,amsmath,amssymb]{revtex4-1}
\usepackage{graphicx}% Include figure files
\usepackage{dcolumn}% Align table columns on decimal point
\usepackage{bm}% bold math

\usepackage{CJK}

\newcommand{\proton}{\ensuremath{p}}
\newcommand{\neutron}{\ensuremath{n}}
\newcommand{\deuteron}{\ensuremath{d}}
\newcommand{\pg}{(\proton,~\ensuremath{\gamma})}
\newcommand{\ag}{(\ensuremath{\alpha},~\ensuremath{\gamma})}
\newcommand{\alal}{(\ensuremath{\alpha},~\ensuremath{\alpha})}

\newcommand{\pd}{(\proton,~\deuteron)}

\newcommand{\alphap}{(\ensuremath{\alpha},~\proton)}
\newcommand{\palpha}{(\proton,~\ensuremath{\alpha})}
\newcommand{\pa}{(\proton,~\ensuremath{\alpha})}

\newcommand{\lis}{\ensuremath{^{7}}Li}
\newcommand{\bes}{\ensuremath{^{7}}Be}

\newcommand{\bte}{\ensuremath{^{10}}B}
\newcommand{\bel}{\ensuremath{^{11}}B}
\newcommand{\cel}{\ensuremath{^{11}}C}
\newcommand{\ctw}{\ensuremath{^{12}}C}

\newcommand{\bag}{\bes\ag\cel}

\newcommand{\helion}{\ensuremath{^3}He}
\newcommand{\hefour}{\ensuremath{^4}He}

\newcommand{\ecm}{\ensuremath{E_\textrm{cm}}}
\newcommand{\thetacm}{\ensuremath{\theta_\textrm{cm}}}
\newcommand{\thetalab}{\ensuremath{\theta_\textrm{lab}}}
\newcommand{\eex}{\ensuremath{E_\textrm{ex}}}
\newcommand{\jpi}{\ensuremath{J^\pi}}
\newcommand{\alphazero}{\ensuremath{\alpha_0}}
\newcommand{\alphaone}{\ensuremath{\alpha_1}}
\newcommand{\pzero}{\ensuremath{p_0}}
\newcommand{\pone}{\ensuremath{p_1}}
\newcommand{\gammaaz}{\ensuremath{\Gamma_{\alpha0}}}
\newcommand{\gammaao}{\ensuremath{\Gamma_{\alpha1}}}
\newcommand{\gammapz}{\ensuremath{\Gamma_{p0}}}
\newcommand{\gammapo}{\ensuremath{\Gamma_{p1}}}
\newcommand{\gammatot}{\ensuremath{\Gamma_\textrm{tot}}}

\begin{document}

%UNCOMMENT
\begin{CJK*}{}{}

%\preprint{APS/123-QED}

\title{Alpha-resonance structure in \cel\
studied via resonant scattering of \bes+$\alpha$ and \bes\alphap\ reaction}

%H. Yamaguchi, T. Hashimoto, S. Hayakawa, D.N. Binh, D. Kahl, S. Kubono, 
%Y. Wakabayashi{$^{\mbox{a}}$}, T.~Kawabata{$^{\mbox{b}}$}, and T.~Teranishi{$^{\mbox{c}}$}\\

\author{H.~Yamaguchi
%UNCOMMENT
}
\email{yamag@cns.s.u-tokyo.ac.jp}
\author{D.~Kahl}

\affiliation{%
Center for Nuclear Study, the University of Tokyo, RIKEN Campus, 2-1 Hirosawa, Wako, Saitama 351-0198, Japan
}%

\author{Y.~Wakabayashi
%UNCOMMENT
}
\author{S.~Kubono
%UNCOMMENT
}
  \affiliation{%
The Institute of Physical and Chemical Research (RIKEN), 
2-1 Hirosawa, Wako, Saitama 351-0198, Japan
}

\author{T.~Hashimoto
%UNCOMMENT
}
  \affiliation{%
Research Center for Nuclear Physics, Osaka University,
10-1 Mihogaoka,Ibaraki, Osaka 567-0047, Japan
}

\author{S.~Hayakawa
%UNCOMMENT
}
  \affiliation{%
Laboratori Nazionali del Sud, Istituto Nazionale di Fisica Nucleare,
Via S. Sofia 62, 95125 Catania, Italy
%Laboratori Nazionali del Sud, via S. Sofia 62, 95125 Catania, Italy
}

\author{T.~Kawabata
%UNCOMMENT
}
\affiliation{%
Department of Physics, Kyoto University, Kita-Shirakawa, Kyoto 606-8502, Japan
}

\author{N.~Iwasa
%UNCOMMENT
}
\affiliation{%
Department of Physics, Tohoku University, Aoba, Sendai, Miyagi 980-8578, Japan
}

\author{T.~Teranishi
%UNCOMMENT
}
  \affiliation{%
Department of Physics, Kyushu University, 6-10-1 Hakozaki, Fukuoka 812-8581, Japan
}

\author{Y.K.~Kwon
%UNCOMMENT
}
  \affiliation{%
Institute for Basic Science, 70, Yuseong-daero 1689-gil, Yuseong-gu, Daejeon 305-811, Korea
}

%\author{J. Hu, S.W. Xu}
%  \affiliation{%
%Institute of Modern Physics, CAS, Nanchang Road 363, 730000 Lanzhou,
%People's Republic of China
%}

\author{D.N.~Binh, L.H.~Khiem, and N.N.~Duy}
\affiliation{%
Institute of Physics, Vietnam Academy of Science and Technology, 18 Hong Quoc Viet, Nghia do, HaNoi, Vietnam
}

\date{\today}% It is always \today, today,

\begin{abstract}
\begin{description}
\item[Background]
The resonance structure in \cel\ is particularly of interest 
with regard to the astrophysical \bes\ag\ reaction, 
relevant at high temperature,
and to the $\alpha$-cluster structure in \cel.
\item[Purpose]
The measurement was to determine unknown
resonance parameters for the high excited states of \cel.
In particular, the $\alpha$ decay width can be useful information
to discuss $\alpha$ cluster structure in \cel.
\item[Methods]
New measurements of the \bes+$\alpha$ resonant scattering and the 
\bes\alphap\bte\ reaction 
in inverse kinematics were performed for center-of-mass energy up to 5.5 MeV, and 
the resonances at excitation energies of 8.9--12.7 MeV in 
the compound \cel\ nucleus were studied. 
Inelastic scattering of \bes+$\alpha$ and the \bes($\alpha$,~\proton$_1$)\bte$^*$ reaction were also studied 
with a  simultaneous $\gamma$-ray measurement.
The measurements were performed at the low-energy RI beam facility CRIB (CNS Radioactive Ion Beam separator)
of the Center for Nuclear Study (CNS), the University of Tokyo. 
\item[Results]
We obtained excitation functions of \bes($\alpha$,~$\alpha_0$)\bes\ (elastic scattering), 
\bes($\alpha$,~$\alpha_1$)\bes$^*$ (inelastic scattering), 
\bes($\alpha$,~\proton$_0$)\bte, and  \bes($\alpha$,~\proton$_1$)\bte$^*$.
Many resonances including a new one were observed 
and their parameters were determined by an R-matrix analysis. 
\item[Conclusions]
The resonances we observed possibly enhance the \bes\ag\ reaction rate 
but in a smaller magnitude than the lower-lying resonances.
A new negative-parity cluster band, similar to the one
previously suggested in the mirror nucleus \bel,  is proposed. 
\end{description}
%% New measurements of the \bes+$\alpha$ resonant scattering and the 
%% \bes\alphap\bte\ reaction 
%% in inverse kinematics were performed for center-of-mass energy up to 5.5 MeV, and 
%% the resonances at excitation energies of 8.9--12.7 MeV in 
%% the compound \cel\ nucleus were studied. 
%% Inelastic scattering of \bes+$\alpha$ and the \bes\alphap\bte$^*$ reaction were also studied 
%% with a  simultaneous $\gamma$-ray measurement.
%% The resonant information in the present range of excitation energy
%% are particularly of interest with regard to the astrophysical \bes\ag\ reaction
%% and alpha-cluster structure in \cel.
%% The measurements were performed at the low-energy RI beam facility CRIB (CNS Radioactive Ion Beam separator)
%% of the Center for Nuclear Study (CNS), the University of Tokyo. 

%The nuclear structure of \bel\ was studied from the \bes+$\alpha$ channel
%at 180$^\circ$  in the center-of-mass system was successfully measured for the first time
%with the inverse kinematics method,
%providing important information on the 
%$\alpha$ cluster structure in \bel\ and the reaction rate of
%\lis\ag, which is relevant to the \bel\ production in the $\nu$-process in core-collapse supernovae. 
%The excitation function of the \lis\alphap\ reaction cross section
%for 11.7--13.1 MeV was also measured.
\end{abstract}


%% 25.40.Ny, Resonance reactions, nucleon-induced
% 24.30.-v Resonances in nuclear reactions, 
% 21.60.Gx, Cluster model, nuclear structure
% Alpha-particle-induced nuclear reactions, 25.55.-e



\maketitle

%UNCOMMENT
\end{CJK*}

\section{Introduction}
%%%%%%%%%%%%%%%%%%555

%A measurement of the \bes+$\alpha$ elastic scattering
%was performed at CRIB \cite{Kubono:02,Yanagisawa:05},
%to study the resonance structure of \cel.
%The excited states of \cel\ above the  
%%\bes+$\alpha$  threshold are 
%threshold for the $\alpha$-particle decay are
%particularly of interest, as described below. 

%The first is on the astrophysical interest.
The \bag\ reaction is considered to play an important role  
in the hot \proton\proton\ chain and related 
reaction sequences \cite{Wiescher:89}.
Several reaction sequences including the \bag\ reaction should take place 
in some high-temperature environments.
One of those sequences is called \proton\proton-V,\\ 

\bag($\beta^{+} \nu$)$^{11}$B(\proton, 2$\alpha$)$^4$He.\\
\\
\noindent Others are rap (II, III and IV) sequences, \\

\bag\pg$^{12}$N\pg$^{13}$O($\beta^{+} \nu$)$^{13}$N\pg$^{14}$O,\\

\bag\pg$^{12}$N($\beta^{+} \nu$)\ctw\pg$^{13}$N\pg$^{14}$O,\\

\noindent and

\bag\alphap$^{14}$N\pg$^{15}$O,\\

\noindent which are reaction chains to synthesize CNO nuclei %$^{14}$O and $^{15}$O.
without the triple-$\alpha$ process, effective 
at $T_9>0.2$, where $T_9$ is the temperature in GK.
At $T_9$ below 0.2, \bes\ predominantly makes an electron capture,
almost independent of the density.
The \bag\ reaction and these sequences possibly play
important roles in the  explosion of supermassive objects with lower metalicity
\cite{Fuller:86}, novae \cite{Hernanz:96} and big-bang nucleosynthesis
\cite{Thomas:94,Coc:12}.
The \bag\ reaction rate is greatly affected by the resonances.
At the lowest temperature, there is a large contribution to 
the reaction rate by the
subthreshold resonance at the excitation energy $\eex=7.50$ MeV. 
%beside the direct capture rate.
The two resonances located at $\eex=8.11$ MeV and $\eex=8.42$ MeV 
determine the rate at higher temperature around $T_9=0.5$--1.
Higher excited states may contribute to the reaction rates
at $T_9>1$.
%The measurement of \bes+$\alpha$ 
%can be associated with the \bes\ag\ reaction.
%The \bes\ag\ reaction is considered to play an important role  
%in the hot \proton-\proton\ chain and related 
%reaction sequences \cite{Wiescher:89}.
%Several reaction sequences including the \bes\ag\ reaction should take place 
%in some high-temperature environments
%($T_{9} > 0.2$).
A recent theoretical calculation \cite{Wanajo:11}
of the $\nu p$-process in core-collapse 
supernovae \cite{Frohlich:06}
shows that 
the \bes\ag\ reaction may contribute 
as much as the triple-$\alpha$ process
to the synthesis of elements heavier than boron
at the relevant temperature of $T_9=1.5$--3.
Therefore, it is also important to study the resonances 
for such a high temperature region, corresponding to $\eex\sim8$--10 MeV
in the Gamow energy window.
%above \eex=
%By our study, the resonant reaction rate should be 
%evaluated more precisely by 
%determining $\alpha$ widths for the resonances 
%corresponding 

The information on the excited states in \cel\ is still limited.
Resonance states above $\eex=9$ MeV have been studied via \bte\pa\ and
other reactions such as \ctw\pd\cel\
\cite{Hunt:57,Brown:51,Chadwick:56,Cronin:56,Jenkin:64,Paul:67,Allan:56,
Overley:62,Olness:65,Hauser:86}.
The resonances typically have widths of the order of 100 keV,
but their $\alpha$-decay widths $\Gamma_\alpha$ are not known with a good precision, 
and even the spin and parity (\jpi) 
have not been clearly determined yet for some of the states.
The excited states at lower energies ($\eex=8$--9 MeV) 
have narrower particle widths, and 
$\Gamma_\alpha$ are only known for 
two resonances located at $\eex=8.11$ and 8.42 MeV.  
The \bag\ reaction cross section was directly measured 
only at the energies of these two resonances \cite{Hardie:84}.
%where the resonance parameters including $\alpha$ widths  
%were determined.
%That measurement is the only known 
%direct measurement of the \bag\ reaction. 

Another interest for the \bes+$\alpha$  system is its nuclear cluster structure.
The 3/2$^{-}_{3}$ state in \cel\ at $\eex=8.11$ MeV is regarded as a
dilute cluster state similar to the one in \ctw\ \cite{Tohsaki:01}, 
where two $\alpha$ particles and \helion\ 
are weakly interacting and spatially extended. %much developed.
Its exotic structure is attracting much attention \cite{Enyo:07}. 
%As mentioned above, this state is also 
%important for the \bes\ag\ reaction rate in stellar sites.
%The situation is similar to the 
%dilute 
%cluster state in $^{12}$C, 
%which plays an important role in 
%astrophysics (the Hoyle state). 
%The Hoyle state, which locates at an
%excitation energy higher than the 3$\alpha$-decay threshold by 0.39
%MeV, is widely considered to have a spatially well-developed 
%3$\alpha$-cluster structure, and extensive theoretical and experimental
%works were devoted to clarify its structure for many decades. 
%Nowadays, this state is considered to have a dilute cluster-gas-like
%structure where three $\alpha$ particles are weakly interacting and are
%condensed into the lowest $s$ orbit \cite{Tohsaki:01}.
The cluster structure in \bel, the mirror nucleus of \cel, was studied
by measuring its isoscalar monopole and quadrupole strengths in the
\bel$(d, d')$  reaction \cite{Kawabata:04,Kawabata:07}. 
As a result, they indicated that the 8.56-MeV state in \bel,
the mirror of the 8.11-MeV state in \cel, 
is considered to have a dilute cluster structure.
%break-up \cite{Soic:04} measurements 
Besides this state, 
rotational bands in \bel\ and \cel, which might be related to the $\alpha$ 
cluster structure, have been discussed \cite{Ragnarsson:81,Soic:04}.
In our recent study of \bel\ by \lis+$\alpha$ resonant elastic scattering, 
we observed strong $\alpha$ resonances, and we determined their $\Gamma_\alpha$.
A new cluster band with negative parity 
was also suggested in the highly excited states \cite{Yamaguchi:11}.

%A similar measurement can be performed for the mirror system, bes+$\alpha$

In the present study, we performed a measurement of 
the \bes+$\alpha$ resonant elastic scattering
to study resonance structure of \cel, 
complementary to the previous study \cite{Yamaguchi:11}.
We also measured protons from \bes\alphap\bte\ reactions,
which have been studied mostly by its inverse reaction.
The actual measurements were performed in inverse kinematics, 
\hefour(\bes,~$\alpha$) and \hefour(\bes,~\proton),
but we denote these as \bes\alal\ and \bes\alphap\bte\
%mainly discussing in the center-of-mass frame, 
for consistency.
% resonant elastic scattering
%to observe $\alpha$ resonances.
The strengths of the resonances in the present study  
should provide useful information on the $\alpha$-cluster structure of \cel\
and on the astrophysical \bes\ag\ reaction rate.
Recently, a similar measurement independently planned 
at other facilities was carried out and 
published by Freer {\it et al.} \cite{Freer:12}.
We will present our new results and 
discuss the differences of the two new measurements.
An essential difference is that we have measured $\gamma$-rays 
to obtain excitation functions of \bes($\alpha$,~$\alpha_1$)\bes$^*$ 
and \bes($\alpha$,~$\proton_1$)\bte$^*$ reactions, which was 
not considered in \cite{Freer:12}.

\section{Method}

\begin{figure}[htbp]
\centerline{\includegraphics[scale=1]{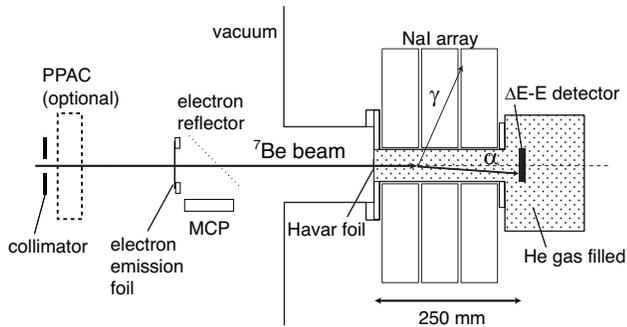}}
\caption{\label{fig:exp_setup} 
Experimental setup of the measurement of the \bes+$\alpha$ elastic scattering
and others in inverse kinematics. 
}
\end{figure}

The measurement %of the \bes+$\alpha$ elastic scattering
was performed at the low-energy RI beam facility CRIB \cite{Kubono:02,Yanagisawa:05},
using the thick target method in inverse kinematics \cite{Artemov:90}
to obtain excitation functions of elastic scattering and others
for $\eex=8.7$--13.0 MeV in \cel.
The experimental setup is almost identical to the 
one used in our \lis+$\alpha$ measurement
\cite{Yamaguchi:11}, except that the beam was an RI beam produced at CRIB.
%\subsection{Beam production}
A pure and intense \bes\ beam can be produced in-flight at CRIB 
using a cryogenic gas target \cite{Yamaguchi:08}.
In the present measurement, a \bes\ beam was produced  
using a 2.3-mg/cm$^2$-thick hydrogen gas target
and a \lis\ beam at 5.0 MeV/u accelerated with an AVF cyclotron.
Having an 8.5-$\mu$m Havar foil as an energy degrader after 
the beam-production target,
a low-energy \bes\ beam at 17.9 MeV was produced.
The \bes\ beam was separated and purified by magnetic analysis and velocity 
selection with a Wien filter.
The purity of the \bes\ beam was about 30\% and almost 100\%
before and after the Wien filter, respectively.
The experimental setup after the Wien filter is shown in Fig.~\ref{fig:exp_setup}.
%Using inverse kinematics, 
%the excitation function at 180 degree in the center-of system, 
%where potential scattering is minimum and the
%resonances can be observed most clearly,
%was measured for the first time.
%The \lis\ beam was accelerated at the AVF cyclotron and  
%transported to the final focal plane (F3) of CRIB.
The beam collimator was 
a 20 $\times$ 20-mm$^2$ rectangular aperture,
accepting a large fraction of 
the transported beam. % transported to the final focal plane (F3) of CRIB.

A Micro-Channel Plate (MCP) was used for the detection of the beam 
position and timing.
A  CsI-deposited 0.7-$\mu$m-thick aluminum foil was placed on the 
beam axis for the secondary electron emission.
The secondary electrons were accelerated along the beam axis
and reflected by 90$^\circ$
at a biased thin-wire reflector 
and detected at the MCP with a two-dimensional delay-line readout. 

The gas target consisted of a 50-mm-diameter duct 
and a subsequent small chamber.
Helium gas at 815 Torr was filled in the chamber and
sealed with a 2.5-$\mu$m-thick Havar foil 
as the beam entrance window.
The \bes\ beam energy after the entrance window of the 
helium gas target target was measured as (15.43$\pm$0.13) MeV.
$\alpha$ particles recoiling to the forward angles
were detected by the $\Delta E$-$E$ detector.
The detector, consisting of 
20-$\mu$m- and 490-$\mu$m-thick silicon detectors,
was placed in the gas chamber.
The helium gas was sufficiently thick to stop the 
\bes\ beam in it before reaching the $\Delta E$-$E$ detector.
The distance from the beam entrance window to the 
detector was 250 mm.
Each detector had an active area of 50 $\times$ 50 mm,
and 16 strips for one side, 
making pixels of 3 $\times$ 3 mm$^2$. 
These detectors were calibrated with $\alpha$ sources,
as well as with 
proton and $\alpha$ beams at various energies produced 
during the run. Each detector had an energy resolution better than 1.5\%
in full width at half maximum (FWHM)
for 5-MeV $\alpha$ particles.
The measurement using the proton and $\alpha$ beams 
was also for the
evaluation of the dead-layer thickness between the two detectors.
%``$\Delta$E'' counter and two or three thick ``E'' counters,
%each with an area of 50 $\times$ 50 mm.
To measure 429-keV $\gamma$ rays from inelastic scattering
to the first excited state of \bes,
Ten NaI(Tl) detectors were placed around the duct and
each NaI(Tl) crystal had
a geometry of 50 $\times$ 50 $\times$ 100 mm$^3$.
They covered 20--60\% of the total solid angle,
depending on the reaction position in the long target.
The energy-dependent photopeak efficiency of the NaI array
was measured at various positions in the gas target,
using standard radioactive  %$\gamma$-ray 
sources of $^{137}$Cs, $^{22}$Na and $^{60}$Co.
The efficiency was determined as 15--30\% for 429-keV $\gamma$ rays. %, depending on the target position.
The energy resolution was about 9\% in FWHM
against 662-keV $\gamma$ rays.

Most of the particles measured at the $\Delta E$-$E$ detector 
were $\alpha$ particles 
from the elastic scattering and protons from the \bes\alphap\ reaction.
Some are in coincidence with $\gamma$ rays, as shown later.
%A small number deuterons were also observed.
%They are unlikely to be produced by a direct reaction at the helium 
%target, but can be produced via at the production target or somewhere 
%% $\alpha$ particles can be produced via the break-up of \lis.
%at the $\Delta E$-$E$ detector,
% in coincidence with the \bes\ beam at the MCP.
%Most of the particles measured was $\alpha$ andfrom the elastic scattering. 
%and a small number of protons and deutrons 
%possibly from \bes\ap\ reaction and break up of \lis\
%were observed in the measurement. 
%when the target gas was replaced to argon 
The typical \bes\ beam  intensity used in the measurement 
was 2 $\times$ 10$^5$ particles per second
at the secondary target, and 
the main measurement using the helium-gas target 
was performed for 4 days, injecting 
2.9 $\times 10^{10}$ \bes\ particles into the gas target.
We performed another measurement using an argon-gas target 
of an equivalent thickness
for 1 day to evaluate the background $\alpha$ particles 
reaching to the the $\Delta E$-$E$ detector 
as a contamination in the secondary beam.

To obtain a correct cross section 
with the current thick-gas target method, 
one needs a correct reaction position
which is determined by the 
geometrical information of the target and detector, 
and energy loss of the beam and $\alpha$ particles.
A measurement with a parallel-plate avalanche counter (PPAC) \cite{Kumagai:01}
in addition to the MCP was also performed for a short time
to check if the cross section is consistent 
between two measurements with different reaction positions for the same \ecm.
In that measurement, the \bes\ beam energy after the entrance 
window was degraded to 12.2 MeV, due to the additional energy loss in the PPAC.
As a result, the reaction position is shifted by 6 cm at maximum 
to the upstream direction, 
which also makes the solid angle smaller, 
as compared with the measurement with the MCP alone.
We confirmed the cross sections finally obtained for both measurements 
were in a good agreement, 
showing that there is no large error in the determination of 
the reaction position.

\section{Deduction of excitation functions}

\subsection{Particle identification}

\begin{figure*}[!htbp]
\centerline{\includegraphics{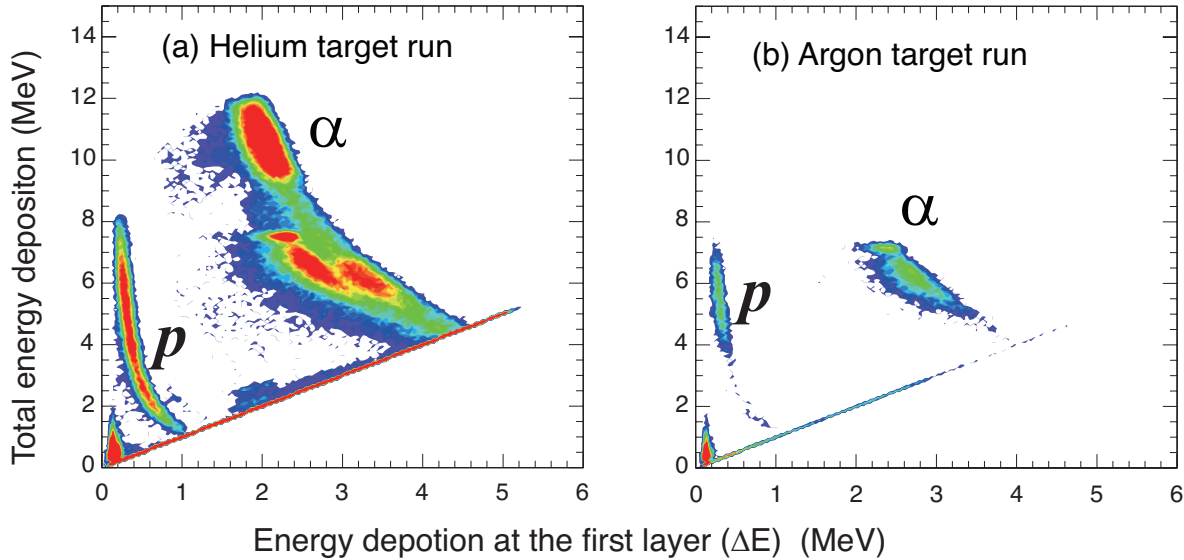}}
\caption{\label{fig:pi} 
(Color online) $\Delta E$-$E$ plot for the particle identification.
}
\end{figure*}

The particle identification performed with the {$\Delta E$-$E$} detector
for the helium and argon target measurements are 
shown separately as two-dimensional energy plots in Fig.~\ref{fig:pi}, where
the total energy deposition of particles is plotted against
the energy deposition in the $\Delta E$ counter.
As illustrated,
$\alpha$ particles and protons were clearly separated.
In the background run with argon gas, we observed non-negligible numbers of 
protons and $\alpha$ particles as shown in Fig.~\ref{fig:pi} (b).
%We expected almost no $\alpha$-particle events for argon-target runs
%as in the case of \lis+$\alpha$ experiment,
%because of the high Coulomb barrier.
%However, we observed 
%events  Figure~\ref{fig:bg_plot}.
They are considered as beam-like 
particles produced at the production target
as contaminants in the \bes\ beam,
and reached the silicon detectors at the end of the beamline.
Most of such particles should have been eliminated at the Wien filter, 
but a very small number comparable to the reaction products 
remained, possibly by scattering in the inner wall of the beamline
or some other reason.

In Fig.~\ref{fig:pi} (a), one may notice the locus of the 
$\alpha$ particles is branched at the total energy about 6 MeV.
This branching is possibly attributed to the channeling effect 
\cite{Alexandrov:92,Poggi:96} of the thin silicon detector.
In the background run, shown in Fig.~\ref{fig:pi} (b), only the left branch 
was prominent, which implies  the background $\alpha$ particles 
from upstream mainly comprise the left branch.
We observed events in the background run also at the right branch,
but the number was much smaller than the left branch.
Such $\alpha$ particles had a strict limitation on the incoming angle 
which was almost perpendicular to the surface of the silicon detector
and thus the channeling effect could be most prominent.
On the other hand, $\alpha$ particles from real scattering events had 
a broader range of angles and are 
less susceptible to the channeling effect.
Although there was such a branching, we successfully
evaluated the amount of background events 
by the argon-target measurement, to be subtracted from
the data of the helium-target measurement.
Performing the argon-target measurement was useful 
not only for the background evaluation, but also for 
understanding the unexpected origin of branching 
in the $\alpha$ particles.

\subsection{Calculation of cross section}

The energy of the \bes\ beam at any position in the gas target 
was obtained with a good center-of-mass energy \ecm\ 
resolution (70 keV or better), based on   
a direct energy measurement at 7 different target pressures,
compared with an energy loss calculation using the SRIM \cite{Ziegler:08} code.
%The reaction position in the target was also determined in 
%the kinematics calculation.
Using this energy loss function,
the energy of recoiled $\alpha$ particle %$E_{\alpha}$ 
measured by the $\Delta E$-$E$ detector was converted to \ecm, 
calculating kinematical relationship for the elastic scattering.
We selected events with an $\alpha$ particle in coincidence with 
a \bes\ beam particle measured in a 20 $\times$ 20 mm$^2$ square 
at the center of the MCP.
%within a 20 $\times$ 20 mm$^2$ square.
The scattering angle determined from the detection position of the $\Delta E$-$E$ detector
was used in the kinematical calculation as a correction.
The energy loss of the recoiled $\alpha$ particle in the gas target
was calculated by SRIM and also considered in the calculation.
The differential cross section $d\sigma/d\Omega$ was calculated for each small energy division using 
the solid angle of the detector, number of beam particles, and the effective target thickness.
The solid angle was calculated by the geometrical information of the detector 
and the reaction position in the target determined by the kinematical calculation. 
The number of beam particles was obtained based on the single counting of the 
beam particle by the MCP, simultaneously recorded in the measurement.
%both of which depend on the reaction position.
Note that it is very important to obtain a correct 
energy loss function in the target, since 
the target thickness and 
the solid angle, both directly reflected in the cross section, 
were also calculated using the function.

In a similar procedure but selecting proton events
and using the kinematical relationship of the \bes\alphap\bte\ reaction,
we obtained a spectrum containing \bes\alphap\bte\ reaction events.

\begin{figure}[!htbp]
\centerline{\includegraphics{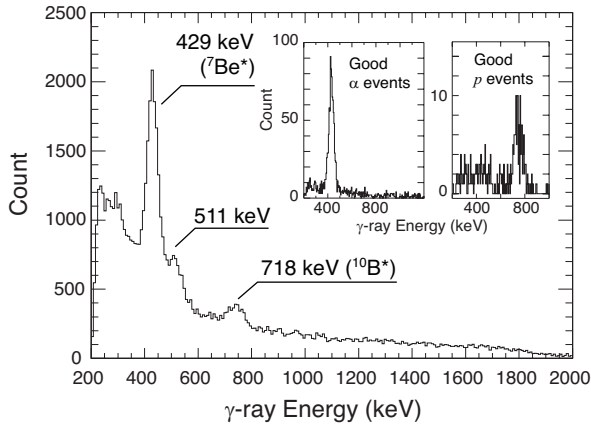}}
\caption{\label{fig:nai_spectrum} 
Energy spectrum of $\gamma$ rays for \bes-$\alpha$-$\gamma$ triple-coincidence events.
The inserts show the same spectrum for events  
with an $\alpha$ particle or \proton\ identified, after a beam position selection.
}
\end{figure}

Events of inelastic scattering \bes($\alpha$,~$\alpha_1$)\bes$^*$ producing 
\bes$^*$ at the first excited state were identified by measuring 
429-keV $\gamma$ rays with the NaI array.
%Considerable numbers of 429-keV $\gamma$ rays were
%detected by accidental coincidence.
We selected triple-coincidence events in which
a \bes\ beam particle was detected at the beam detector (MCP),
an $\alpha$ or any other particle at the silicon detectors and a $\gamma$-ray 
at the NaI array.
The $\gamma$-ray energy spectrum for those events is shown in Fig.~\ref{fig:nai_spectrum}.
The peak at 429 keV was clearly identified, and
small peaks around 511 keV and 718 keV,
which should be from positron annihilations and
excited $^{10}$B$^*$ produced via the \bes($\alpha$,~\proton$_1$)\bte$^*$ reaction
 respectively, were also observed.
We performed a finer event discrimination by taking events 
in which an $\alpha$ particle was identified
and the beam is hitting at the central part of the MCP,
an energy spectrum of $\gamma$-rays  with 
a good signal-to-noise ratio was obtained,  
as shown in the insert of Fig.~\ref{fig:nai_spectrum}.
The events with a 429-keV  $\gamma$-ray were used to obtain 
the excitation function of the inelastic scattering.
A similar event discrimination was successfully 
performed for protons as also shown in the insert of the figure, 
to select events of the \bes($\alpha$,~\proton$_1$)\bte$^*$ reaction.
%Although the statistics does not appear to be very good.
The photopeak efficiency of the NaI array, 
measured at various position in the 
gas target, was used for the calculation of the 
absolute cross section.
%By kinematical calculations and the normalizations by the the efficiency, 
Finally,
excitation functions of the inelastic scattering and 
the \bes($\alpha$,~\proton$_1$)\bte$^*$ reaction were obtained. 
%The small contribution by the accidental coincident events 
%was evaluated from the total excitation function and subtracted.

\subsection{Background Subtraction}

\begin{figure}[!htbp]
\centerline{\includegraphics{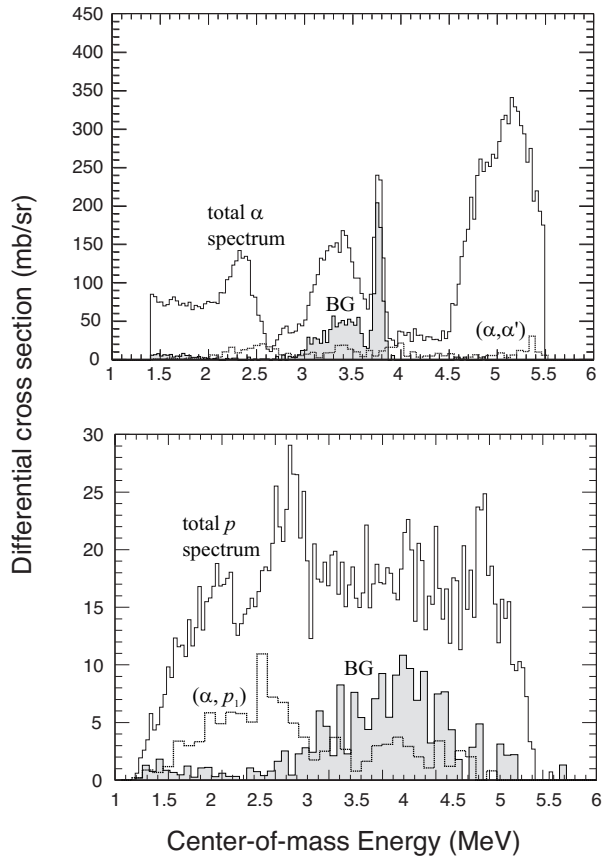}}
\caption{\label{fig:bg_plot} 
Background subtraction for $\alpha$ (upper) and \proton\ (lower) spectra.
Total $\alpha$ and \proton\ spectra are 
the excitation functions of \bes\alal\bes\ and
\bes\alphap\bte, but still containing background
events.
Shaded spectra show the contributions of 
background $\alpha$ and \proton\ as beam contaminations.
The spectra of 
\bes($\alpha$,~$\alpha_1$)\bes$^*$
and \bes($\alpha$,~\proton$_1$)\bte$^*$ are also shown.
}
\end{figure}

The \bes\alal\bes\ and  \bes\alphap\bte\ spectra obtained by the above procedure 
still contain background
$\alpha$ and \proton\ contributions,
%as beam contaminations, 
the amount of which could be evaluated by 
the argon-target measurement.
%Measurements were also performed using an argon target 
%having almost equivalent 
%stopping power against the \bes\ beam.
The background contribution to the differential cross sections 
were evaluated  as shown in Fig.~\ref{fig:bg_plot}.

The sharp peak at 3.7 MeV in the background $\alpha$ spectrum 
corresponds to $\alpha$ particles 
which had the magnetic rigidity 
analyzed at the dipole magnets (D1 and D2) in CRIB.
%coming with a ....trajectories.
The broader lower-energy component is possibly from 
$\alpha$ particles which had the same origin,
but were scattered somewhere in the beamline.
%These beam-like particles were rejected to some extent
%by the Wien filter and by an event discrimination
%based on the timing information, but still remained in the obtained spectra.
The heights of the sharp peak for both target runs are in good agreement,
but the broad component was significantly higher in the helium-target spectrum.
This suggests the peak around 3-3.5 MeV observed in the helium-target spectrum is 
partially due to the background $\alpha$, but the rest is by real
scattering events.
%Thus, the background contribution evaluated by the argon-target run data 
%was subtracted from the helium-target spectrum.
By subtracting the background contribution and 
the contribution of inelastic scattering events,
we obtained an excitation function 
of elastic scattering.

\begin{figure*}[!htbp]
\centerline{\includegraphics{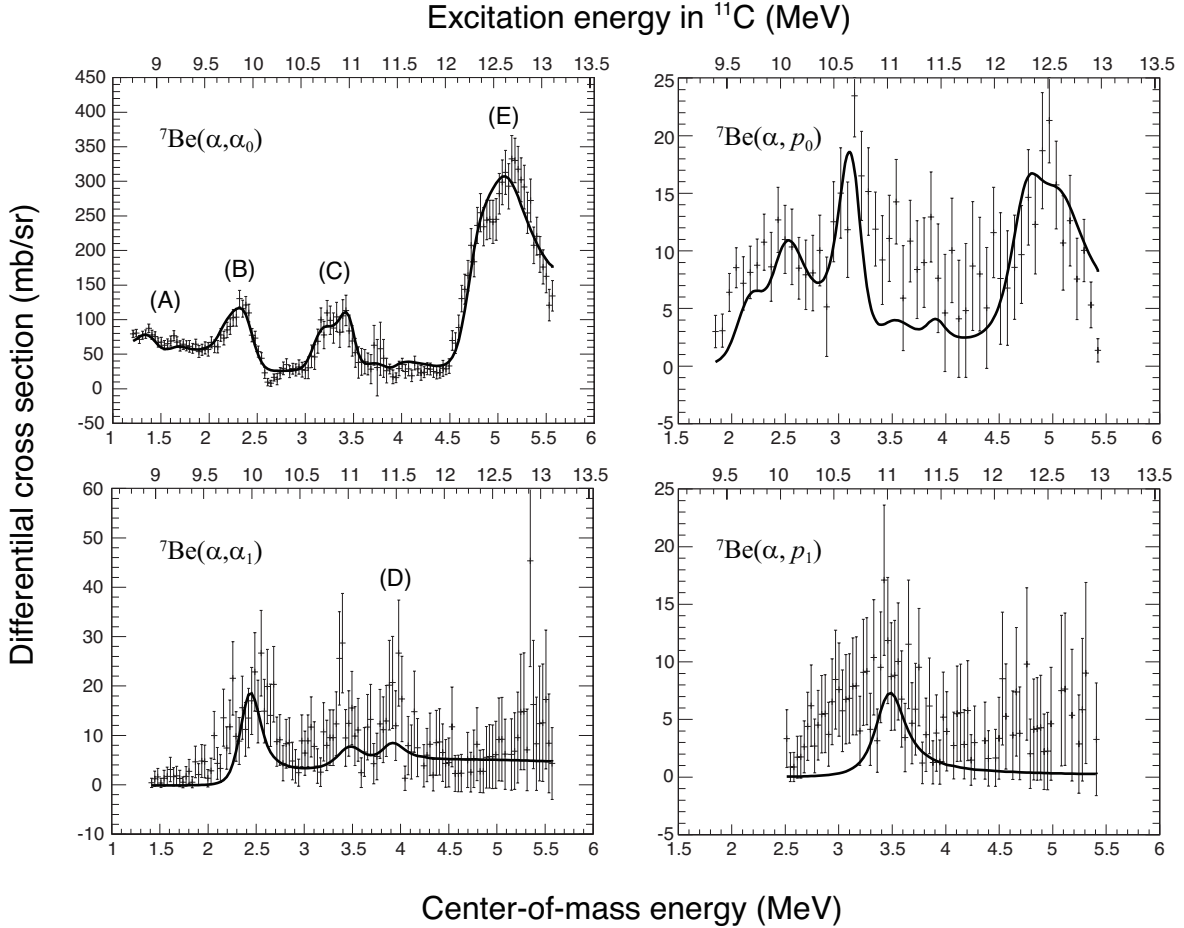}}
\caption{\label{fig:ex_functions} 
Excitation functions, referred to as 
\alphazero, \alphaone, \pzero, and \pone\ spectra.
The best-fit curves by the R-matrix analysis are also shown.
The labels (A--E) correspond to structures discussed in Sec.~\ref{sec:r-matrix}.
}
\end{figure*}

There was no sharp peak in the background proton spectrum, but 
we performed a similar subtraction for the broad background
and obtained the \bes\alphap\bte\ spectrum.
The contribution from the \bes($\alpha$,~\proton$_1$)\bte$^*$ 
reaction was also considered.
Finally we obtained excitation functions for four different cross sections,
\bes($\alpha$,~$\alpha_0$)\bes, \bes($\alpha$,~$\alpha_1$)\bes$^*$, 
\bes($\alpha$,~\proton$_0$)\bte, and  \bes($\alpha$,~\proton$_1$)\bte$^*$. 
We refer to these excitation functions as the \alphazero, \alphaone, \pzero, and \pone\ spectra, 
respectively, in the following sections.
The excitation functions are shown in Fig.~\ref{fig:ex_functions}.
The peak structure observed in the excitation functions
should correspond to the resonances in \cel.
The fitted curves are by an R-matrix analysis, which will be described later.

%where  \bes$^*$ or \bes$^*$ denotes the first excited state of each nucleus.

\subsection{Energy and angular uncertainty}

\begin{figure}[!htbp]
\centerline{\includegraphics{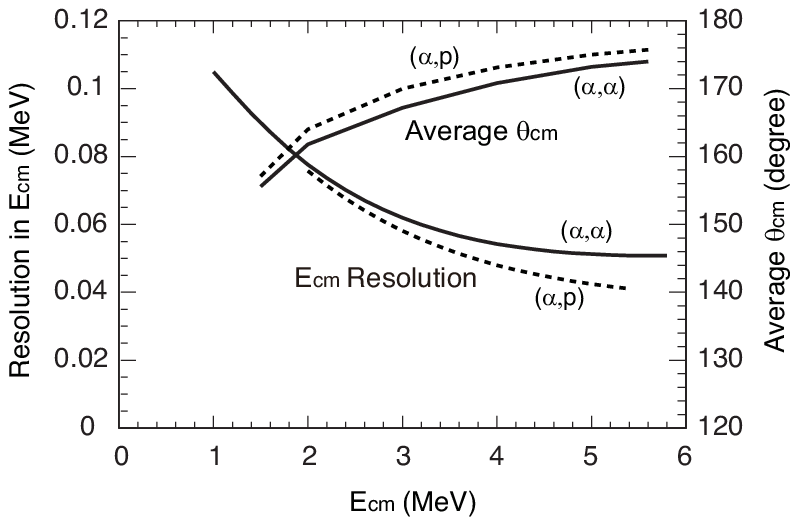}}
\caption{\label{fig:resolution} 
Uncertainty in \ecm\ and averaged \thetacm\
for the \alphazero\ and \pzero\ spectra.
}
\end{figure}

Uncertainty in \ecm\ and averaged center-of-mass scattering angle \thetacm\ 
are plotted for the \alphazero\ and \pzero\ spectra in Fig.~\ref{fig:resolution}, and 
the curves for  \alphaone\ and \pone\ spectra are quite similar to those.

For the \alphazero\ spectrum, the overall uncertainty in \ecm\ was estimated as
70--130 keV, depending on the energy.
The uncertainty mainly originated from 
the energy straggling of the \bes\ beam and $\alpha$ particles (30--80 keV),
the energy resolution of the $\Delta E$-$E$ detector (20--55 keV),
and the angular uncertainty due to the finite size of the detector (20--100 keV).
The uncertainty is similar for the \alphaone\ spectrum, and 
slightly better for the \pzero\ and \pone\ spectra.

The excitation functions in Fig.~\ref{fig:ex_functions} are for 
certain angular range covering $\thetacm=180^\circ$.
The average \thetacm\ was 167$^\circ$ at $\ecm=3$ MeV,
but it depends on \ecm.
%It should be noted that the averaged angle is not flat against \ecm,
%especially at lower \ecm.
The dependence is because the $\Delta E$-$E$ detector 
and the long helium gas target were closely arranged
to obtain good statistics, and we had to select events within   
a fixed area of the detector in the analysis.

%The present excitation function was for \thetacm\ 160--180$^\circ$, 
%and 

%The statistics of inelastic events was not sufficient
%to discuss the detailed structure, but 
%the differential cross section 
%was about 30 mb/sr at maximum, one order of magnitude smaller than 
%that of the elastic scattering.
%smaller than the one for elastic scattering.
%Such a small cross section is in agreement
%with the measurements of the mirror system \cite{Yamaguchi:11}.

%% In summary,
%% we successfully obtained excitation functions of
%% \bes+$\alpha$ elastic scattering and inelastic scattering
%% to the first excited state of \bes.
%% %The cross section of the inelastic scattering 
%% %was much less than that of the elastic scattering.
%% We will perform an R-matrix analysis 
%% to determine resonance parameters,
%% which are not known precisely from previous measurements. 
%% The $\alpha$ widths would provide 
%% valuable information for the $\alpha$-cluster structure 
%% in the high excited states of \cel, and 
%% astrophysical \bes\ag\  reaction rate in high-temperature phenomena.

\section{Comparison with previous results}

\subsection{Comparison with previous inverse reaction measurements}
A direct comparison is possible for our \pzero\ spectrum
with the previous measurements of the inverse  
\bte\palpha\bes\ reaction \cite{Cronin:56,Jenkin:64,Overley:62},
although the measured scattering angles are not the same.
The laboratory cross sections of the previous measurements were converted 
into the center-of-mass cross section, using the detailed balance theorem. 
%Data in \cite{Jenkin:64} with \thetalab\ at 90$^\circ$, and 150$^\circ$ 
%(corresponding \thetacm\ = 59$^\circ$, 102$^\circ$, 156$^\circ$ for $E_p=4 MeV$), 
%\cite{Cronin:56} at 90$^\circ$, and \cite{Overley:62} at 90$^\circ$ are shown.
\begin{figure*}[!htbp]
\centerline{\includegraphics{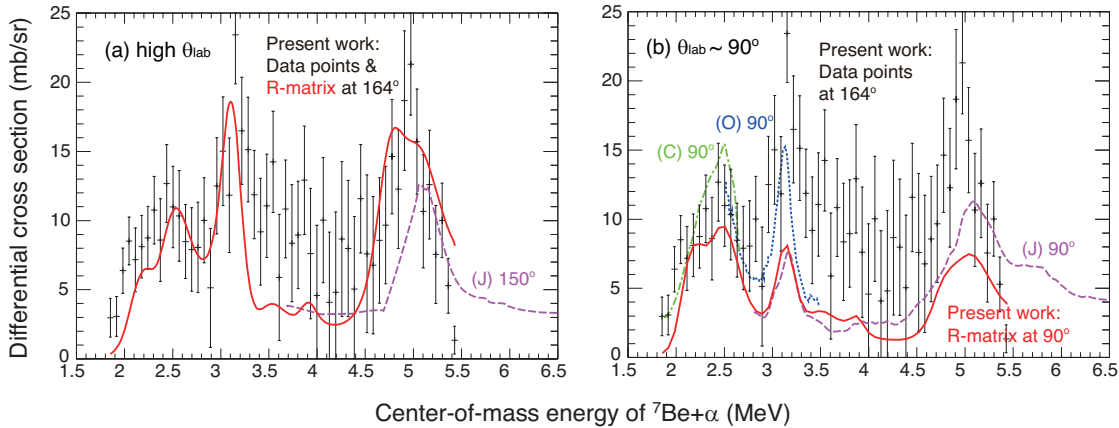}}
\caption{\label{fig:inverse} 
(Color online) The \pzero\ spectrum compared with 
previous measurements of the inverse \bte\palpha\bes\ reaction by 
(J) Jenkin {\it et al.} \cite{Jenkin:64},
(C) Cronin {\it et al.} \cite{Cronin:56},
and (O) Overley {\it et al.} \cite{Overley:62}.
R-matrix calculations performed at $\thetalab=90^\circ$ and 164$^\circ$ are also shown.
Note that the angles are in laboratory system of the \bte\palpha\bes\ reaction.
}
\end{figure*}
The left panel in Fig.~\ref{fig:inverse}(a) shows the data in \cite{Jenkin:64} with 
the laboratory angle $\thetalab=150^\circ$ and 
the present measurement with an R-matrix calculation.
%of which the details discussed later. 
The present measurement had a corresponding average angle 
in the inverse reaction of $\thetalab=164^\circ$ (at $\ecm=3$ MeV). 
The right panel (b) shows the data in \cite{Cronin:56,Jenkin:64,Overley:62}, all with 
$\thetalab=90^\circ$, but the present measurement data shown are the same as in (a).
The curve for the present work is by an R-matrix calculation performed 
with the same resonant parameters as in (a), but with the angle adjusted 
to these previous measurements.

%The present data have a large uncertainty, 
%because of the low statistics and systematic error by the subtraction procedure.
Considering the large uncertainty of our measurement,
the overall features of the present spectrum  are in agreement with 
previous measurements, such that the two peaks around 3 and 5 MeV are distinct.
The absolute cross section shows a disagreement to some extent, but
the previous measurements already had differences  
with one another as in Fig.~\ref{fig:inverse}(b).
The cross section by \cite{Cronin:56} or \cite{Overley:62} show
a similar magnitude of the cross section with ours,
but the angle is at $\thetalab=90^\circ$.
The cross section by \cite{Jenkin:64} at the same angle is lower than 
these two and our data at $\thetalab=164^\circ$.
The angular dependence may partly explain the difference,
as shown by the R-matrix curves at two different angles.
One can see the calculation at $\thetalab=90^\circ$ is
closer to the data by \cite{Jenkin:64} at $\thetalab=90^\circ$.
%although the  R-matrix calculation may not 
%accurately reproduce such angular dependence in general.
The cross section of the present work at 3.5--4.5 MeV 
appears higher than any of previous data and the R-matrix calculation. 
We did not introduce strong resonances in this region to improve the fitting, 
since such resonances were not observed in the previous measurements.
This discrepancy might be related to the background protons,
which were distributed around 3--4.5 MeV, as shown in Fig.~\ref{fig:bg_plot}. 

\subsection{Comparison with the latest inverse kinematics measurement}
\begin{figure}[!htbp]
\centerline{\includegraphics{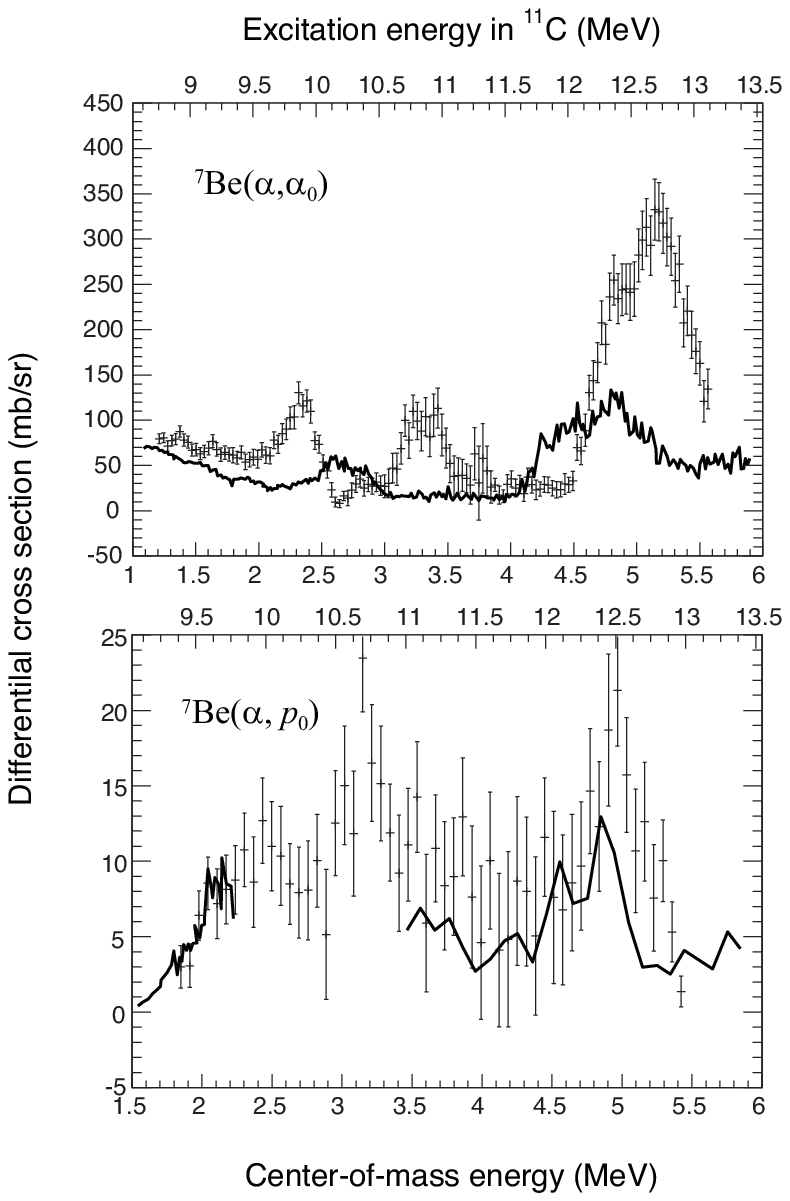}}
\caption{\label{fig:compare_freer} 
The \alphazero\ and \pzero\ spectra by 
the present work and \cite{Freer:12} (solid lines). 
}
\end{figure}

The measurement by Freer {\it et al.} \cite{Freer:12} has been performed 
with essentially the same method as the present work.
Here we compare the two results.
The \alphazero\ and \pzero\ spectra of both measurements 
are compared in Fig.~\ref{fig:compare_freer}.
The overall shapes of the two \alphazero\ spectra have a common feature.
There is a large double-peak structure around $\ecm=5$ MeV,
and smaller peaks are located at an energy 1 MeV below, and at lower energies.
However, two large differences are seen;
the absolute cross section is different by a factor 2--3 for 
the higher energy part,
and the peaks are displaced in energy by about 500 keV.
The cross section in \cite{Freer:12} was normalized 
using the low-energy measurement data, and also by  
a Monte Carlo simulation, which may produce
a systematic uncertainty as much as factor 2.5 at maximum, as they claim.
On the other hand, our cross section is based on single counting of the beam particles 
and geometrical measurement.
Furthermore, the previous \lis+$\alpha$
measurement \cite{Yamaguchi:11} with the same analysis method
yielded a consistent cross section with 
older normal kinematics measurements. 
Therefore, an error in the cross section 
by a factor 2--3 is unlikely to be produced in our measurement or analysis.
%Nevertheless, the around 5 MeV
%strong peaks are consistent, 
%as shown later.
In the thick target method, the energy can be easily shifted 
if the energy loss functions of the \bes\ beam or the $\alpha$ particle 
in the target is not correct.
In our both measurements, the energy loss of the beam was directly measured 
over a wide range of pressures.
We used the SRIM code for the energy loss of the $\alpha$ particle, and
the higher edge of \ecm\ from the measured data (5.5 MeV)
was in good agreement 
with the \ecm\ expected from the 
measured beam energy at the beginning of the target.
We do not expect an error in \ecm\ 
much more than 100 keV at \ecm $\sim$ 5 MeV.
Another fact that may support the present work is 
that our spectrum could be explained with the known energy level information,
as shown later, but in \cite{Freer:12} it was necessary 
to introduce several new resonances to perform an R-matrix fit.

The \pzero\ spectrum in \cite{Freer:12} was separated into two 
energy regions.
The lowest energy part perfectly agree with ours, but 
the data in \cite{Freer:12} at higher energy again shows 
a lower cross section. However, the discrepancy is not obvious, 
since the deviation is comparable to 
the uncertainty, quite large in both measurements.

\section{R-matrix analysis}
\label{sec:r-matrix}

Several resonance structures were observed in our \alphazero\ spectrum,
namely, (A) Two small peaks at at 8.9--9.2 MeV, 
(B) a peak around 10 MeV, (C) a broad peak at 10.5--11.0 MeV,
considered to be a doublet, and (E) a  doublet structure 
at 12--13 MeV, as shown in Fig.~\ref{fig:ex_functions}.
Structures were also observed
in the other three spectra, and 
some are corresponding to the peaks in the \alphazero\ spectrum, 
and others are not, 
as (D) a small enhancement of the cross section in the \alphaone\ spectrum
at 11.4 MeV.

We performed an analysis using 
an R-matrix calculation code (SAMMY8 \cite{SAMMY:00})
to deduce resonance parameters.
The decay widths of four channels, \gammaaz, \gammaao, 
\gammapz, and \gammapo, were considered in the R-matrix analysis.
The basic strategy was as follows.
First \gammaaz\ were roughly determined by an R-matrix fit of the \alphazero\ spectrum,
reproducing the peak structure.
Using the \gammaaz, we analyzed the \pzero\ spectrum 
and found the best values for  \gammaaz\ and \gammapz.
Then, the \alphazero\ spectrum was analyzed again to determine \gammaaz\ more precisely.
When \gammaaz\ and \gammapz\ were determined consistently by the two spectra, 
analysis was performed for the \alphaone\ and \pone\ spectra,
both of which had low statistics and do not exhibit much structure.
By repeating the above process until it converges, 
all the four widths, \gammaaz, \gammaao, 
\gammapz, and \gammapo, were determined.
%The statistics are small for the last two spectra, and 
%a strong restriction was not made for  \gammaao\ and \gammapo\.
To cope with many parameters in the fitting, we adopted
resonant parameters by previous measurements \cite{Ajzenberg:90} as much as possible.
The detailed discussions for each structure we observed (A--E), including the 
\jpi\ determination, are below.
Another restriction we considered was that
the sum of the decay widths do not exceed the known 
total width \gammatot\ too much.
The calculation was performed with channel radii $R_c=4.2$ fm for
the \bes+$\alpha$ channel, and $R_c=3.8$ fm for the \bte+\proton\ channel.
These channel radii were deviated by 20\% for the evaluation of the uncertainties. 
The angle was fixed at $\thetacm=167^\circ$ for the \alphazero\ and \alphaone\ spectra,
and  $\thetacm=169^\circ$ for the \pzero\ and \pone. 

The results are summarized in Table~\ref{tab:resonance_params}. 
\begin{table*}[!htbp]
\caption{\label{tab:resonance_params} Best-fit resonance parameters of \bel\
determined by the present work. 
The \eex\ and $J^{\pi}$ shown in italic letters
were fixed to those in \cite{Ajzenberg:90,TUNL:04}, and 
the others are proposed in the present work.
See text for other possible \jpi\ assignments.}
\begin{ruledtabular}
\begin{tabular}{ccccccccc}
\eex\  & $J^{\pi}$ &$l_{\alpha0}$ & $\Gamma_{\alpha0}$  &$\Gamma_{\proton0}$  & 
     $\Gamma_{\alpha1}$  &$\Gamma_{\proton1}$  &
     $\Gamma_{\mathrm{tot}}$ \cite{TUNL:04} & $\Gamma_{\mathrm{W\alpha}}$\\ 
(MeV)  & & &  (keV) &(keV) &(keV) &(keV) &(keV) &(keV)\\
\colrule
8.90        & (9/2$^{+}$)          & 3& 8         &    &    &    &   &6.4 \\ 
{\it 9.20}  & {\it   {5/2}$^{+}$  }& 3& 13        &    &    &    &500&21  \\
{\it 9.65}  & {\it  ({3/2}$^{-}$) }& 0& 20        & 50 &    &    &210&1310\\
{\it 9.78}  & {\it  ({5/2}$^{-}$) }& 2& 19        &100 &    &    &240&450 \\
{\it 9.97}  & {\it  ({7/2}$^{-}$) }& 2&153$\pm$55 & 35 & 30 &    &120&580 \\
{\it 10.083}& {\it    7/2$^{+}$   }& 3& 25        &230 &    &    &230&90 \\ 
{\it 10.679}& {\it    9/2$^{+}$   }& 3& 58$\pm$36 &110 &    &    &200&230 \\ 
{\it 11.03} & (5/2$^{-}$)         & 3&130$\pm$83 & 25 &45  &120 &300&360 \\
{\it 11.44} & (3/2$^{+}$)         & 1& 80        & 30 &150 &    &360&2680\\
{\it 12.40} & 9/2$^{+}$           & 3&460$\pm$150& 90 &    &    &1000--2000&1100\\
{\it 12.65} & {\it  (7/2$^{+}$) } & 3&420$\pm$178&110 &    &    &360&1270\\
%    12.63 $\pm$ 0.04  & (3/2$^{+}$\cite{Ajzenberg:90}, 5/2$^{+}$, 7/2$^{+}$, 9/2$^{+}$\cite{Soic:04})\footnotemark
% \footnotetext{3/2$^{+}$ and 9/2$^{+}$ were suggested in previous studies, 
%while four spins are possible from our measurements alone. }
\end{tabular}
\end{ruledtabular}
\end{table*}
The Wigner limit %$\Gamma_w=2 \hbar/R_n (2E/\mu)^{1/2} \mu P_l$ 
$\Gamma_w=2 \hbar^2/\mu R^2 P_l$,
where $\mu$ is the reduced mass and $P_l$ is the penetrability, 
was calculated for an $\alpha$ particle with an interaction radius $R=4.2$ fm 
and shown in the table for comparison.
%and $\gamma_\alpha^2$ was calculated by $\Gamma_\alpha/2P$, where $P$
%is the penetration factor.
Uncertainties could be reasonably evaluated only for some \gammaaz, 
as shown in Table~\ref{tab:resonance_params}, and 
the other widths could have very large uncertainties.

\subsection{Peaks at 8.9--9.2 MeV}
In this energy region the excitation function was rather flat 
in the \alphazero\ spectrum,
but two small bumps were observed.
One of these may correspond to the known 5/2$^+$ state
at 9.20 MeV \cite{Wiescher:83}.
The other one is located around 8.90 MeV.
A resonance at this energy is not known by previous measurements
and we regard this as a new resonance.
However, it can be the same resonance as the one
at 8.655 (7/2$^+$) or 8.699 MeV (5/2$^+$) \cite{TUNL:04},
because the energy uncertainty 
in this lowest energy region is quite large.

The 9.20-MeV resonance was initially introduced by 
Wiescher {\it et al.} \cite{Wiescher:83}
in their analysis of the \bte\pg\ reaction measurement.
%This resonance is not observed in any other previous measurements.
Although a large total width of $\gammatot=500$ keV was incorporated in the analysis 
of Ref.~\cite{Wiescher:83},
our calculation with a large \gammapz\ resulted in 
diminishing the peak height, and 
a large \gammaaz\ far exceeding the Wigner limit becomes 
necessary.
In the best fit, \gammapz\ was assumed to be 0. 

For the 8.90-MeV bump, a resonance with $l_\alpha=3$ gives a good fit,
and \jpi=9/2$^+$ was the best among them.
Other possible \jpi\ were 3/2$^+$, 5/2$^+$, and 7/2$^+$.
The resonance in this energy region may enhance the
astrophysical \bes\ag\ reaction rate, as shown later.
However, it is difficult to derive conclusive resonance parameters
with such small peaks broadened by the energy resolution,
averaged for a broad range of angles. 
The current R-matrix calculation could be deceptive at the edge of our energy range.
Therefore, it is desirable to perform another study for the resonances 
in this energy region.

%\subsection{Resonances at 9.6--10.1 MeV}
\subsection{Peak around 10 MeV}

The \alphazero\ spectrum exhibits a peak,
which could be well reproduced by a 
resonance around 10.0 MeV having a large \gammaaz.

Four resonances at 9.65, 9.78, 9.97 and 10.083 MeV 
are known in this region \cite{Ajzenberg:90}.
%\jpi\ were tentatively assigned as 3/2$^-$, 5/2$^-$, 7/2$^-$
%for the first three, and 
The 9.78-MeV and 10.083-MeV resonances have been observed by many experiments.
The former was by the 
\bte\pg\ \cite{Hunt:57,Day:54} and
\bte\pa\ reactions \cite{Hunt:57,Brown:51,Chadwick:56,Cronin:56,Jenkin:64,Paul:67,Allan:56},
%Cronin...isotropic distribution...3/2- is the favored. (5/2+,7/2+)
%by the mirror level in 11B.
and the latter was by the 
\bte\pa\ \cite{Hunt:57,Brown:51,Chadwick:56,Cronin:56,Allan:56}
and \bte(\deuteron,~\neutron) reactions \cite{Overley:62}.
%Brown observes anomalies in (p,p) data also at 9.78 and 10.084.
%\bte\pa\bes\ \cite{Ophel:62,Bernstein:64}
However, the \jpi\ was in controversy 
for the former state \cite{Brown:51,Overley:62},
while firmly determined as 7/2$^+$ for the latter one.  

The current \jpi\ assignments in this region were
mostly by Wiescher {\it et al.} \cite{Wiescher:83}.
They made tentative assignments of \jpi\ 
for the three states at 9.65, 9.78 and 9.97 MeV 
as 3/2$^-$, 5/2$^-$,  and 7/2$^-$, respectively, 
by an analysis of the \bte\pg\ reaction measurement data.
The 9.65 and 9.97 MeV states, which had not been known before,
were introduced to reproduce their excitation functions.
Later, a resonance near 9.97 MeV was also
observed by the $^{12}$C\pd\ reaction \cite{Smith:84}, but 
\jpi\ was not determined.
No other observation of the 9.65-MeV resonance is known.
The 9.65-MeV (3/2$^-$), 9.78-MeV (5/2$^-$), and 10.083-MeV (7/2$^+$) 
states have candidates of 
their mirrors at 10.26, 10.34, 10.60 MeV in \bel,
while the 9.97-MeV state has none.

We fully adopted the tentative \jpi\ assignments
in \cite{Wiescher:83}.
%The \alphazero\ spectrum exhibits a peak,
%which could be well reproduced by a 
%resonance at 9.97 MeV having a large \gammaaz.
The absence of this resonance in the present \pzero\ spectrum
and previous \bte\alphap\ measurements suggests 
this 9.97-MeV resonance should have a much smaller \gammapz\
than the neighboring 9.78-MeV and 10.083-MeV resonances.
There is a peak around 9.97 MeV also in the \alphaone\ spectrum,
and we introduced \gammaao\ in the calculation to reproduce the peak.

%\subsection{10.679  MeV}
\subsection{Doublet at 10.5--11.0 MeV}

A peak was observed in the \pzero\ spectrum,
as previously observed by the inverse reaction \cite{Cronin:56,Jenkin:64,Overley:62}. 
There is a peak in the \alphazero\ spectrum at the same \ecm, 
and the spectral shape suggests that 
it may be forming a doublet with another resonance at 11.0 MeV.
%In the present \alphazero\ spectrum, this resonance should correspond to 
For the higher component in the doublet,
a good fit was obtained by a 
$l_\alpha=2$ resonance with $\jpi=5/2^-$ or  $7/2^-$. 

A resonance at 11.03 MeV had been observed by $^{11}$B(\helion, $t$) \cite{Watson:71}, 
$^{13}$C(\proton, $t$) \cite{Benenson:74}, and others \cite{TUNL:04}.
%21,23,28
%10B(p,\alpha)
%12C(p,d)11C
%12C(3He,a)11C
However, \jpi\ has not been determined, and  
only assumed as a state with an isospin $T=1/2$.

In the \pone\ spectrum,
a resonance was observed around 11.03 MeV
as the only peak structure in the spectrum.
This resonant shape could be reproduced by introducing a resonance with 
$\jpi=$(3/2$^-$, 5/2$^-$, 7/2$^-$), but the peak height was 
significantly lower, when \jpi\ was assumed to be 3/2$^-$.
There is a small peak-like structure also in the \alphaone\ spectrum.
The peak is consistent with a calculation introducing a $\jpi=5/2^-$ resonance,
although the agreement in the peak height is not obvious
due to the large uncertainty.
In a calculation with $\jpi= 7/2^-$,
$l_{\alpha1}=4$ was required for the \bes+\alphaone\ channel,
and a good fit was not obtained.
Here we adopt \jpi=5/2$^-$ as the best assignment,
but $\jpi=7/2^-$ might be also possible,
if we ignore the \alphaone\ spectrum which has quite low statistics.

\subsection{Structure at 11.4  MeV}

A small enhancement of the cross section was 
identified at 11.4 MeV in the \alphaone\ spectrum.
%11.44 MeV...Jenkin 
A resonance at 11.44 MeV was observed by Jenkin {\it et al.} \cite{Jenkin:64} 
by the \bte(\proton,~$\alpha_1$)\bes$^*$ reaction 
only at forward scattering angles, but the resonance was not 
clearly seen by the \bte(\proton,~$\alpha_0$)\bes\ reaction.
Their cross section of the \bte(\proton,~$\alpha_1$)\bes$^*$ reaction 
was simply decreasing as \thetalab\ increases. 
The implication of this angular dependence was not discussed in detail.
An R-matrix calculation with a single resonance cannot reproduce
such an angular dependence asymmetric with regard to $\thetacm=90^\circ$.
However, a resonance with a spin
$J$=3/2--7/2 can have a feature that the cross section 
decreases at backward angles toward $\thetacm=180^\circ$.

In the present \pzero\ spectrum, we did not clearly observe 
the resonance either.
This suggests the resonance should have a relatively large \gammaao.
Considering the absence of a sharp peak in the \alphazero\ and \pzero\ spectrum
and the best R-matrix fitting for the \alphaone\ spectrum,
we tentatively assigned \jpi\ as 3/2$^+$, but
$\jpi=3/2^-$, 5/2$^\pm$, and 7/2$^\pm$ might be possible.
Note that this resonance in the \alphazero\ spectrum
is closely located to the sharp peak in the background contribution
as shown in Fig.~\ref{fig:bg_plot}, and a small structure could be lost. 
%The \alphaone\ channel has a \jpi=1/2$^-$, and
%the large width...

%Such feature can be 
%The strong angular dependence suggests that the spin is higher than 1/2,
%but the width 

%high spins are excluded because 
%requires angular momentum to form such a resonance from 
%7Be+a1 channel (\jpi=1/2-), and 
%the width is limited.

\subsection{Doublet at 12--13  MeV}

In this energy region, we observed a double-peak structure in the
$\alpha_0$ spectrum with a large cross section.

Ignoring isospin $T=3/2$ states, resonances at 12.4 and 12.65 MeV have been
known previously. % in this energy region.
%Jenkin 7/2+ p 0.20 a0 0.15 a1 0.05 MeV in Lab.
In a previous \bte\palpha\bes\ reaction measurement \cite{Jenkin:64}, a broad resonance 
having a large width of 400 keV was observed. 
Assuming a single level at 12.65 MeV, they
tentatively determined \jpi\ as 7/2$^+$.
The 12.4-MeV resonance was observed 
by the \bte\pg\ reaction \cite{Kuan:70}, 
and by the \ctw(\helion,~$\alpha$) reaction.
In \cite{Kuan:70}, the width was determined as  1--2 MeV and 
it was discussed as a part of a giant resonance.

In the present work, the observed double-peak structure 
was fitted by the 7/2$^+$ resonance at 12.65 MeV, and 
another resonance at 12.4 MeV.
The best fit was made with $\jpi=9/2^+$ for the 12.4-MeV resonance.
The fit was unsatisfactory by any other \jpi,
although it was considered as a negative parity state in \cite{Kuan:70}.
Freer {\it et al.} \cite{Freer:12} also explained
this structure as a doublet consisting of 9/2$^+$ and 7/2$^+$ resonances,
although the doublet is displaced in energy by several 100 keV, and 
the ordering is reversed.
%and the obtained \gammapo\ and \gammaao\ are consistent
%with our present work.
The broad peak we observed in the \pzero\ spectrum at the same energy 
was also fitted by the doublet of the same two states.

\section{Discussion}

\subsection{Alpha-cluster bands}

\begin{figure}[!htbp]
\centerline{\includegraphics{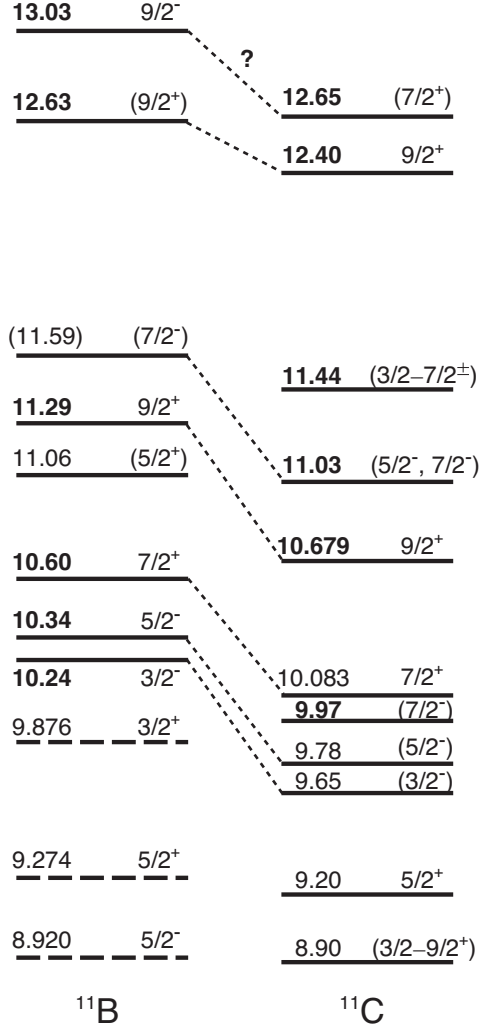}}
\caption{\label{fig:mirror_level} 
Resonant states observed in the present work and our
previous work \cite{Yamaguchi:11}.
\eex\ in MeV and \jpi\ in our works are shown for each state, and 
the states with an \eex\ in bold letters were observed as significant resonant peaks.
The states with dashed lines were not observed 
in our measurements, but taken from \cite{TUNL:04}.
}
\end{figure}

The strong resonances we have observed in the \alphazero\ spectrum
have large \gammaaz, which are reflecting their property 
of $\alpha$-cluster structure.
Fig.~\ref{fig:mirror_level} shows resonant states in \bel\ and \cel\
observed in the present work and our previous work \cite{Yamaguchi:11}.
We can identify several pairs of mirror states, as indicated in the figure.
The difference in \eex\ is about 500 keV for the lower pairs of states and 
smaller for the highest two.
Such energy difference between mirror states was discussed in \cite{Freer:12} 
in relation with the phase transition from shell-model states to cluster states 
\cite{Itagaki:05,Itagaki:08}.

Rotational bands in \bel\ and \cel, 
which might be related to the cluster structure, 
had been discussed in \cite{Ragnarsson:81,Soic:04}.
In our previous work \cite{Yamaguchi:11},
we have indicated that 
the 12.63 MeV resonance in \bel\ may have $\jpi=9/2^+$,
as initially mentioned in \cite{Soic:04}.
%% and belongs to the $K=$3/2$^+$ band.
We also proposed a new negative-parity band,
of which the head is the 8.56-MeV ($\jpi=3/2^-$) state.
According to a recent calculation based on 
antisymmetrized molecular dynamics (AMD) method \cite{Suhara:12},
this can be interpreted as a negative-parity band 
having a large B(E2) of 20--30 $e^2$fm$^4$, and  
the members should have a 2$\alpha$-$t$ cluster structure.
The energy of the band head appeared as 
lower than the line expected from the other members,
in both the experiment and theory.
In \cite{Suhara:12}, the lowering in the level energy was attributed to 
the relatively weak interactions between 2$\alpha$-$t$ in the 8.56-MeV state,
making a deviation from the higher states 
which have a more rigid structure.

\begin{figure}[!htbp]
\centerline{\includegraphics{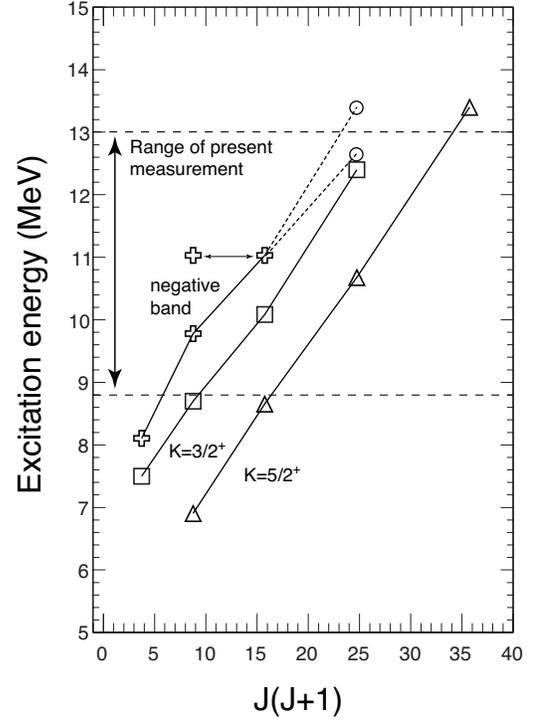}}
\caption{\label{fig:rot_band} 
Two positive-parity rotational bands in \cel\ suggested 
in \cite{Soic:04}, 
and a newly proposed  negative-parity band.
%showing that 
%the excitation energy has a linear dependence on $J(J+1)$.
}
\end{figure}

A similar discussion can be applied to \cel.
Two positive-parity rotational bands, $K=3/2^+$ and 5/2$^+$,
were suggested in \cite{Soic:04}.
We observed a strong resonance with 
$\jpi=9/2^+$ at 12.4 MeV in \cel, and
it can be the missing member of 
the $K=3/2^+$ rotational band. 
We propose a new negative-parity band
 also in \cel,
as shown in Fig.~\ref{fig:rot_band}.
The members of the band could have  
a 2$\alpha$-\helion\ cluster structure.
%It is assumed that 
%the excitation energy has an almost linear dependence on $J(J+1)$, but 
%whether the band is simply rotational or not is unknown 
The head is the 8.10-MeV ($\jpi=3/2^-$) state,
the mirror of the 8.56-MeV state in \bel. 
The second member is the 9.78-MeV state,
assigned as $\jpi=5/2^-$ previously.
The third member can be the state at 11.03 MeV.
Our assignment was either $\jpi=5/2^-$  or 7/2$^-$,
and the latter assignment agrees with the systematics of 
this negative band.  
The systematics predicts there can be a $\jpi=9/2^-$ 
state around 13 MeV.
A candidate is the 12.65-MeV state.
In the present work,
this state was regarded to form a doublet together with the 
12.4-MeV state. 
A similar doublet was observed in the mirror nucleus \cite{Yamaguchi:11},
and the higher state was considered to have $\jpi=9/2^-$ (see Fig.~\ref{fig:mirror_level}). 
If the tentative assignment of $\jpi=7/2^+$ was wrong,
the 12.65-MeV state may have $\jpi=9/2^-$, as in the mirror nucleus. 
Another candidate is the resonance at 13.4 MeV \cite{TUNL:04}, 
which is known to have a certain $\alpha$ width,
but its \jpi\ has not been determined.
In Fig.~\ref{fig:rot_band}, these states are shown as circles 
and connected as dotted lines under the assumption that 
they have $\jpi=9/2^-$. 
The energy of the band head (8.10 MeV) appears 
as lower than the systematics expected from the higher state. 
This lowering is in agreement with the 
mirror state \cite{Yamaguchi:11,Suhara:12}.

A problem is that the \gammaaz\ of the 9.78-MeV (5/2$^-$) resonance 
is not large, while the neighboring 9.97-MeV resonance with $\jpi=7/2^-$
has a larger \gammaaz, 
being more likely to be an $\alpha$-cluster state.
The previous studies in the  mirror nucleus \cite{TUNL:04, Yamaguchi:11} show
there is no such corresponding state with $\jpi=7/2^-$ in \bel, 
as shown in Fig.~\ref{fig:mirror_level}.
In this respect, the current identification of the resonances and the 
\jpi\ assignments for $\eex=9.5$--10 MeV can be questioned.

\subsection{Astrophysical reaction rate of  \bes\ag}

The resonances observed in the present work might contribute to
the astrophysical \bes\ag\cel\ reaction rate at high temperature, $T_9>1.5$. 
Here we calculate the resonant reaction rate and 
compare it with the total reaction rate evaluated in NACRE \cite{NACRE:99,NACRE:11}.
In the evaluation in NACRE, only 2 resonances 
at 8.1045 and 8.420 MeV are included.
These two resonances dominate the reaction rate 
$N_\textrm{A} \langle \sigma v \rangle $  up to the temperature
$T_9\sim 3$,  and a Hauser-Feshbach calculation 
rate was included for the higher temperature. 

\begin{figure}[!htbp]
\centerline{\includegraphics{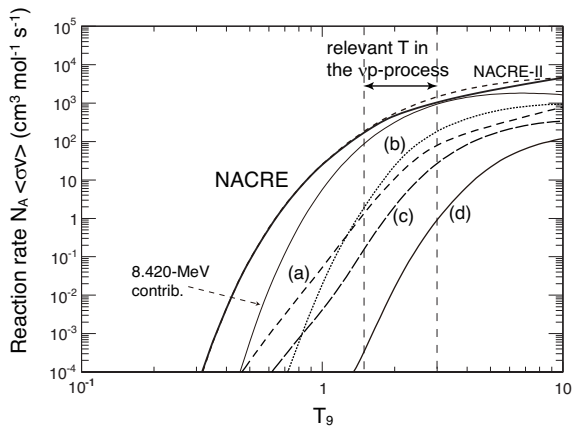}}
\caption{\label{fig:rr} 
Resonant reaction rate of \bes\ag\ for the 8.90, 9.20, and 9.97-MeV resonances,
calculated by the analytical formula 
(see text and Table~\ref{tab:rr_params} for the labels).
The evaluation by NACRE and NACRE-II are 
shown for comparison. The contribution by the 8.420-MeV resonance,
included in NACRE, is also shown.
%$T_9$ is the temperature in GK.
}
\end{figure}

The resonant reaction rates were calculated for three resonances 
using analytical formula described in \cite{NACRE:99}, and
plotted in Fig.~\ref{fig:rr}.
The total reaction rate evaluated by NACRE 
and its recently updated rate (NACRE-II) \cite{NACRE:11}  are 
also shown for comparison.
%% As for the resonance at \eex=12.63 MeV, calculation results for  
%% \jpi\ based on previous studies (3/2$^+$, 9/2$^+$) are shown.
Table~\ref{tab:rr_params} shows the parameters we used in the calculation.
\begin{table*}[!htbp]
\caption{\label{tab:rr_params} 
Parameters used in the reaction rate calculation.
The dominant destination states of the $\gamma$ decay according to our
calculation are also shown.}
\begin{ruledtabular}
\begin{tabular}{rcccccccc}
&\eex\  &$J^{\pi}$ &$\Gamma_{tot}$ (keV)& $\Gamma_\alpha$ (keV) & $\Gamma_\gamma$ (eV) 
& $\omega$ & $\omega \gamma$ (eV) &dominant dest.\\
 \colrule
(a)&8.90 & 9/2$^{+}$ & 8    & 8  & 0.48   & 2.5 & 1.2 & 8.655 MeV (7/2$^+$)\\
(b)&8.90 & 3/2$^{+}$ & 8    & 8  & 1.7    & 1.0 & 1.7 & ground (3/2$^-$)\\
(c)&9.20 & 5/2$^{+}$ & 500  & 13 & 34     & 1.5 & 1.3 & ground (3/2$^-$)\\
(d)&9.97 & 7/2$^{-}$ & 218  &153 & 0.97   & 2.0 & 1.4 & 6.90 MeV (5/2$^+$)\\
\end{tabular}
\end{ruledtabular}
\end{table*}
%% $\Gamma_\alpha$ is known from our measurement. 
The $\Gamma_\gamma$ and the decay scheme are experimentally unknown 
for this energy range. 
Therefore, we evaluated $\Gamma_\gamma$ by 
a calculation based on the Weisskopf unit
with a spectroscopic factor of 0.1, which roughly reproduces the
experimentally known $\Gamma_\gamma$ of the 8.1045 and 8.420-MeV resonances.
Note that such a spectroscopic factor was not explicitly used 
in the plot of our previous publication \cite{Yamaguchi:11}.
%% We used the total widths $\Gamma_{tot}$ in \cite{TUNL11:90},
%% or fixed at our $\Gamma_\alpha$, when it exceeds
%% the total width in \cite{TUNL11:90}.  
%% $\omega$ and $\omega \gamma$ are the statistical factor and
%% the strength of the resonance, respectively.
%The calculated contribution to the reaction rate is shown in Fig.~\ref{fig:rr}
%for each parameter sets, 
The labels (a)--(d) in Fig.~\ref{fig:rr} 
correspond to the ones in Table~\ref{tab:rr_params}.
(a) is for our newly identified resonance at 8.90 MeV.
%under the assumption of assignment of the 
%The data in this energy region had large uncertainties in
%the energy and scattering angle, and 
%the \gammaaz\ and the \jpi\ were not determined with a good confidence.
\jpi\ was taken to be 9/2$^+$ from our best fit.
Since the \jpi\ assignment may not be correct, we 
also evaluated the contribution 
for the same resonance, but when the 
resonance had a lower spin of 3/2, shown as (b).
Basically (b) contributes to the reaction rate more, but  
its tail contribution is smaller than (a).
For $T_9=2$--3, 
where the 8.420-MeV resonance dominates the reaction rate,
the contribution of the 8.90-MeV 
was evaluated as around 10\% of the total 
reaction rate.
(c) is for the 9.20-MeV resonance, 
where we also found a small peak in the \alphazero\ spectrum.
We used $\jpi=5/2^+$ and \gammatot=500 keV from previous measurements, 
although such a large \gammatot\
was inconsistent with our R-matrix analysis.
(d) is for the 9.97-MeV resonance, which was identified as a strong 
alpha resonance in the \alphazero\ spectrum.
The tentative assignment of $\jpi=7/2^-$ \cite{Wiescher:83} was used 
for the evaluation, and 
a smaller contribution was obtained, as shown in Fig.~\ref{fig:rr}.
We also evaluated contributions for higher resonances, but none of 
them had an effect as much as or larger than the case (d).

In summary, the resonances at 8.90 MeV and 9.20 MeV
have a possibility to give significant contributions to the reaction rate 
for $T_9=1.5$--3, 
although they are unlikely to be more than 
the contribution of the 8.420-MeV resonance, which dominates the reaction rate.  
Considering that the $\Gamma_\gamma$ used here 
could be underestimated by factors and the decay widths and \jpi\ are
also uncertain, more studies are favored for the determination of 
resonant parameters in the energy region of $\eex=8.5$--9.5 MeV,
which might be difficult to access from the \bte+\proton\ channel.
On the other hand, the resonances 
above 9.5 MeV can be considered as negligible for $T_9<10$.

%% b) and c) are of the same resonance, but
%% the magnitudes differ. 
%% The reason is that the resonant state b) can decay to the ground state directly
%% with an E1 transition, but c) cannot, because of its high spin.
%% As shown in the plot,
%% the only considerable contribution from the resonance of case b)
%% occurs in the very high temperature region, $T_9>5$.
%% Considering that the Weisskopf unit tends to overestimate the width
%% in such cases, the contribution for the other high spin resonances 
%% are likely to make no significant contribution to the 
%% reaction rate even at $T_9=10$.
%% However, studies on the gamma widths and decay scheme 
%% and determination of \jpi\ are needed for a
%% conclusive evaluation of the reaction rate.

\section{Summary}

%We have studied $\alpha$ resonant scattering of \bes\ 
%and \bes\alphap\ reaction 
%We measured the excitation functions for cross sections of 
%the \bes+$\alpha$ elastic and inelastic scatterings.
We have studied resonant states for $\eex=8.7$--13.0 MeV in \cel\ via 
$\alpha$-resonant elastic scattering with 
the thick-target method in inverse kinematics,
using a low-energy \bes\ beam at CRIB.

We obtained excitation functions of 
\bes($\alpha$,~$\alpha_0$)\bes, \bes($\alpha$,~$\alpha_1$)\bes$^*$, 
\bes($\alpha$,~\proton$_0$)\bte, and  \bes($\alpha$,~\proton$_1$)\bte$^*$
simultaneously, measuring $\gamma$-rays in coincidence with $\alpha$ particle or \proton.
The excitation function
of the elastic scattering exhibited strong $\alpha$ resonances 
mostly in agreement with previous measurements, and we 
brought new information on the resonance parameters
with an R-matrix analysis.
The \bes($\alpha$,~\proton$_0$)\bte\ excitation function was consistent
with the previous measurement of the inverse reaction, \bte\palpha\bes.
The excitation functions were compared with 
the ones by a similar measurement performed recently \cite{Freer:12}, 
and we found differences in the absolute cross section and 
the energy, although their spectral shapes had a similarity.

A new negative parity band, which could have 
2$\alpha$-\helion\ cluster structure,
is proposed in \cel,
in accordance with the previously proposed band in the mirror nucleus \bel.
The resonant contribution for the astrophysical reaction rate of \bes\ag\cel\ was
evaluated at high temperature using the new resonance parameters, 
and a 10\%-order enhancement
over the evaluation by NACRE could be expected for $T_9=1.5$--3.

%% up to \eex=13.1 MeV (\ecm=4.4 MeV), using the thick-target method in inverse kinematics.
%% The excitation function of the elastic scattering exhibited 
%% resonances mostly consistent with previous measurements,
%% and we successfully determined their resonance parameters.
%% In particular, a reliable determination for $\alpha$ decay widths was
%% made for the first time. 
%% A \jpi=1/2$^+$ and $T=$3/2 resonance was known to be at 12.56 MeV, 
%% but we proposed the existence of 
%% another $T=$1/2 resonance at 12.63 MeV,
%% of which \jpi\ is possibly 3/2$^+$ or 9/2$^+$.
%% \bel\ is known to have rotational bands with 
%% a large moment of inertia.
%% We proposed a new negative-parity band 
%% consistent with a theoretical calculation,
%% but its character (e.g. rotational or not)
%% should be studied more in detail in the future.
%% We evaluated the resonant reaction rate of \lis\ag\bel\ 
%% at high temperature using the new $\alpha$ widths, but a major enhancement
%% over the evaluation by NACRE is unexpected for $T_9<5$.\\

\begin{acknowledgments}
The experiment was performed at RI Beam Factory operated
by RIKEN Nishina Center and CNS, the University of Tokyo.
We are grateful to the RIKEN and CNS accelerator staff for their help.
We truly appreciate Prof. Y.K.~En'yo and Mr. T.~Suhara for useful discussions
and suggestions based on their theoretical calculations.
This work was partly supported by JSPS KAKENHI (No. 21340053) and
the Grant-in-Aid for the Global COE Program 
``The Next Generation of Physics, Spun from Universality and Emergence'' 
from the Ministry of Education, Culture, Sports, Science and Technology (MEXT) of Japan. 
One of us (L.H. Khiem) wishes to acknowledge the support by Vietnam National Foundation for 
Science and Technology (NAFOSTED) under Contract No.103.04.54.09.
\end{acknowledgments}

\bibliography{crib}

\end{document}